\documentclass[reprint, amsmath, amssymb, aps, prb, superscriptaddress,floatfix]{revtex4-2}

\usepackage{graphicx}
\usepackage{dcolumn}
\usepackage{amsmath}
\usepackage{amsfonts}
\usepackage{epsfig}
\usepackage{bm}
\usepackage{color}
\usepackage{mathtools}
\usepackage[version=4]{mhchem}

\begin{document}


\title{Sign-Free Determinant Quantum Monte Carlo Study of Excitonic Density Orders in a Two-Orbital Hubbard-Kanamori Model}

\author{Xu-Xin Huang}
\affiliation{Department of Applied Physics, Stanford University, Stanford, CA 94305, USA}
\affiliation{Stanford Institute for Materials and Energy Sciences, SLAC National Accelerator Laboratory and Stanford University, 2575 Sand Hill Road, Menlo Park, CA 94025, USA}
\author{Brian Moritz}
\affiliation{Stanford Institute for Materials and Energy Sciences, SLAC National Accelerator Laboratory and Stanford University, 2575 Sand Hill Road, Menlo Park, CA 94025, USA}
\author{Martin Claassen}
\affiliation{Department of Physics and Astronomy, University of Pennsylvania, Philadelphia, PA 19104, USA}
\author{Thomas P. Devereaux}
\affiliation{Stanford Institute for Materials and Energy Sciences, SLAC National Accelerator Laboratory and Stanford University, 2575 Sand Hill Road, Menlo Park, CA 94025, USA}
\affiliation{Department of Materials Science and Engineering, Stanford University, Stanford, CA 94305, USA}
\affiliation{Geballe Laboratory for Advanced Materials, Stanford University, Stanford, CA 94305, USA}

\date{\today}

\begin{abstract}
While excitonic instabilities in multiorbital systems recently have come under scrutiny in a variety of transition-metal compounds, understanding emergence of these instabilities from strong electronic interactions has remained a challenge. Here, we present a sign-problem-free determinant quantum Monte Carlo study of excitonic density orders in a half-filled two-orbital Hubbard-Kanamori model with broken orbital degeneracy, which accounts for the role of Hund's coupling in transition-metal compounds. For strong inverted (negative) Hund's exchange, we find numerical evidence for the emergence of excitonic density order, with competition between anti-ferro-orbital order and $\mathbf{Q} = (\pi,\pi)$ excitonic density order as a function of orbital splitting and Hund's coupling. While inverted Hund's coupling stabilizes a spin-singlet excitonic density phase for weak orbital splitting, positive Hund's coupling favors a spin-triplet excitonic density phase. 

\end{abstract}

\pacs{}

\maketitle

\section{\label{sec:level1}Introduction}

Harnessing unconventional phases of matter in strongly-correlated electron systems holds great promise for the functionalization of quantum properties of materials. At the same time, the inability to capture the collective behavior of electrons in such systems from simple descriptions of the mean behavior of individual quasiparticles poses a fundamental theoretical challenge, necessitating the development of new and unbiased computational many-body techniques to gain further insight. 
Despite its innocent appearance, the one-band Hubbard model remains challenging for both analytical and numerical studies in two dimensions\cite{arovas2021hubbard,huang2021intertwined}. Even richer physics can emerge in transition-metal compounds with partially-filled $d$ shells, ranging from spin-orbital fluctuations\cite{PhysRevB.80.235114,Kita_2009,PhysRevB.91.195149} and orbital order\cite{PhysRevB.72.205124,PhysRevB.72.205126,PhysRevLett.102.126401,mezio2019orbital} to Hund's coupling-mediated superconductivity\cite{PhysRevB.70.054513,PhysRevB.70.172504,PhysRevB.99.024516}, further exacerbating the theoretical challenge. Gaining insight into emergent unconventional electronic phases using exact numerical simulations of minimal multi-orbital models is hence essential, to provide important clues to understand correlated electron behavior in complex materials.

Excitonic density orders constitute an intriguing possible phase that can emerge in strongly-interacting models with multiple valence orbitals. An exciton is a bosonic bound state of an electron and a hole, which can undergo Bose-Einstein condensation below a critical temperature\cite{lozovik1976new,RevModPhys.90.021001} to form an excitonic superfluid. Extensive early studies have analyzed excitonic condensation and bi-exciton formation in bilayer quantum wells and heterostructures with spatially-separated electrons and holes \cite{doi:10.1063/1.5052674, PhysRevLett.88.206401, PhysRevLett.110.216407, PhysRevB.78.045313, PhysRevB.79.125308, sharma2018phase}. Meanwhile, studies of superconductivity in strongly correlated materials recently have motivated explorations of electron-hole counterparts in multi-layer lattice models\cite{0953-8984-27-33-333201, PhysRevB.88.235115, PhysRevB.88.235127, PhysRevB.85.165135, fujiuchi2018excitonic, PhysRevB.77.144527, PhysRevB.91.144510, PhysRevB.88.035312, 1742-6596-529-1-012030}. For sufficiently strong inter-layer interactions, a previous work also provided evidence for bi-exciton condensation in a two-orbital Hubbard model with density interactions and lifted orbital degeneracy\cite{PhysRevLett.124.077601}.

The exploration of excitonic orders naturally extends to multi-orbital compounds with Hubbard-Kanamori interactions. Here, variational calculations for inter-orbital density interactions have found an induced excitonic instability\cite{PhysRevB.85.165135,PhysRevB.84.144425}. An excitonic instability close to a spin-state transition was identified using dynamical mean-field theory (DMFT) calculations\cite{PhysRevB.89.115134}; the role of Hund's coupling in selecting excitonic pairing symmetries was studied further using the variational cluster approximation \cite{PhysRevB.90.245144}; and the phase diagram and spectral functions of symmetry allowed excitonic pairings and magnetic orders near the excitonic instability were investigated in Ref.~\cite{PhysRevB.90.235140} and Ref.~\cite{PhysRevLett.122.127601}.  

Meanwhile, in the absence of exact numerical results, the propensity for orbital excitonic order for strong electronic interactions remains an open question. Exact diagonalization calculations are limited to very small system sizes \cite{PhysRevB.88.035312}, whereas efforts using determinant quantum Monte Carlo (DQMC) have remained constrained to high temperatures due to the fermion sign problem and the complicated interaction structure of multi-orbital models. DQMC is a numerically unbiased method for simulating interacting quantum many-body systems at finite temperature with polynomial effort in system size, whereby the interacting quantum mechanical problem is replaced via a stochastic sum of single-particle problems, suitable for importance sampling \cite{PhysRevD.24.2278,PhysRevB.40.506,santos2003introduction}. However, a notorious sign problem typically emerges for fermion or frustrated systems, whereby negative sampling weights can lead to large statistical errors. However, the sampling weight sign is a basis dependent property and in principle can be avoided by using symmetries and clever choices for the auxiliary field in DQMC.

In this work, we present sign-problem-free determinant quantum Monte Carlo simulations for the emergence of excitonic density orders in two-orbital models with Hund's interactions and lifted orbital degeneracy. Our results provide unbiased numerical evidence for both singlet and triplet excitonic density orders induced via strong Hund's exchange. To this end, we utilize a new auxiliary-field decomposition scheme for density interactions and Hund's coupling of the Hubbard-Kanamori model, and prove the absence of the fermion sign problem. The numerically-deduced phase diagram is consistent with the intuition yielded by devising simple strong-coupling theories that capture the emergence of excitons at low energies. Combined, our work provides convincing evidence for emergent excitonic density order in two-orbital transition-metal compounds.

\section{Model and Methods}

\subsection{Two-Orbital Hubbard-Kanamori Model}

Consider a minimal model of transition-metal ions with a half-filled $e_g$ shell on a square lattice
\begin{align}
\hat{H}&= \hat{H}_{0}+\sum_{i}\hat{H}_{I}
\end{align}
with
\begin{align}
\hat{H_{0}}&=-t\sum\limits_{\langle i,j\rangle,\alpha,\sigma} \hat{c}^{\dagger}_{i\alpha\sigma}\hat{c}_{j\alpha\sigma} - \Delta\sum\limits_{i\sigma}(\hat{n}_{iA\sigma}-\hat{n}_{iB\sigma}), \\
\hat{H}_{I}&=U(\hat{n}_{A\uparrow}\hat{n}_{A\downarrow}+\hat{n}_{B\uparrow}\hat{n}_{B\downarrow})+U'\sum_{\sigma\sigma'}\hat{n}_{A\sigma}\hat{n}_{B\sigma'} \nonumber \\
    &-J\sum_{\sigma\sigma'}\hat{c}^{\dagger}_{A\sigma}\hat{c}_{A\sigma'}\hat{c}^{\dagger}_{B\sigma'}\hat{c}_{B\sigma} \nonumber \\
    &+J(\hat{c}^{\dagger}_{A\uparrow}\hat{c}^{\dagger}_{A\downarrow}\hat{c}_{B\downarrow}\hat{c}_{B\uparrow}+\hat{c}^{\dagger}_{B\uparrow}\hat{c}^{\dagger}_{B\downarrow}\hat{c}_{A\downarrow}\hat{c}_{A\uparrow}) \nonumber \\
    &-\mu\sum\limits_{\sigma}(\hat{n}_{A\sigma}+\hat{n}_{B\sigma}).
\label{hamiltonian}
\end{align}
Here, $\hat{c}^{\dagger}_{i\alpha\sigma}(\hat{c}_{i\alpha\sigma})$ are creation (annihilation) operators for an electron at site $i$ in orbital $\alpha\in \{A, B\}$ with spin $\sigma\in \{\uparrow, \downarrow\}$ and the number operator $\hat{n}_{i\alpha\sigma}\equiv\hat{c}^{\dagger}_{i\alpha\sigma}\hat{c}_{i\alpha\sigma}$. $\hat{H}_0$ encodes the bilinear kinetic and crystal field terms, whereas $\hat{H}_I$ describes the on-site interactions, where the site index for fermion operators in $\hat{H}_{I}$ has been omitted for simplicity. The parameter $t$ denotes the hopping amplitude between nearest neighbors. Note that we only consider intra-orbital hopping, and the hopping amplitude is the same for orbital A and orbital B. $\Delta$ lifts the orbital degeneracy per site, permitting asymmetric orbital filling. On a single site, $U$ and $U'$ parameterize intra-orbital Hubbard and inter-orbital density interactions, respectively, while $J$ parameterizes the spin- and pair-exchange Hund's interactions between orbitals. In addition, $\mu$ denotes the chemical potential, which controls the overall doping level.

When $J=0\,t$, total charge in orbital A and orbital B is conserved separately, and $\hat{H}_{I}$ in Eq.~(\ref{decomp}) is invariant under a $U(1)\times U(1)$ symmetry. Excitonic condensation (EC) now entails a spontaneous breaking of the $U(1)$ orbital charge symmetry. However, in the absence of $J$, each layer furthermore remains individually invariant under spin rotations. Hence, the resulting excitonic condensate would need to break both $U(1)$ charge and $SU(2) \times SU(2)$ spin rotation symmetry, precluding a finite-temperature phase transition in two dimensions, and favoring instead a Kosterlitz-Thouless transition to an exotic bi-excitonic condensate \cite{PhysRevLett.124.077601} at finite $\Delta$.

In the presence of Hund's exchange ($J\neq 0\,t$), the charge conservation symmetry is reduced to $U(1)\times Z_{2}$: only the total charge is conserved, with the pair hopping term of Hund's exchange leaving intact a residual charge parity symmetry per individual layer. Furthermore, the spin exchange term retains only a global $SU(2)$ spin rotation symmetry. Excitonic order now breaks the \textit{discrete} $Z_2$ symmetry, and will be denoted henceforth as an excitonic density wave (EDW) instability.


Interestingly, in the strong coupling limit, the two-orbital Hubbard-Kanamori model introduced above can be mapped to a transverse-field Ising model for pseudo-spins of excitonic nature. Suppose that $U \sim U'$ and $J/t < 0$ with $|J| \gg t$. Here, the ground state for a single two-orbital site in the strong coupling limit is two-fold degenerate and forms a pseudo-spin 1/2 with 
\begin{align}
|\Uparrow_i\rangle&=\frac{1}{\sqrt{2}}\left( \hat{c}^{\dagger}_{iA\uparrow}\hat{c}^{\dagger}_{iB\downarrow}-\hat{c}^{\dagger}_{iA\downarrow}\hat{c}^{\dagger}_{iB\uparrow}\right)|0\rangle, \\
    |\Downarrow_i\rangle&=\frac{1}{\sqrt{2}}\left( \hat{c}^{\dagger}_{iA\uparrow}\hat{c}^{\dagger}_{iA\downarrow}+\hat{c}^{\dagger}_{iB\uparrow}\hat{c}^{\dagger}_{iB\downarrow}\right)|0\rangle, \
\end{align}
which describes precisely a hardcore boson (spin-singlet exciton).

The role of weak hopping $t$ can now be captured in a strong-coupling expansion in $t/J$, yielding a transverse-field Ising model for the excitonic pseudo-spins
\begin{align}
\label{effH2}
    \hat{H}'=&J_x\sum_{\langle i,j\rangle}\hat{T}^{x}_{i}\hat{T}^{x}_{j}+h_{z}\sum_i \hat{T}^{z}_{i},
\end{align}
where
\begin{align}
    \hat{T}^{+}_i&=|\Uparrow_i\rangle \langle\Downarrow_i|, \\
    \hat{T}^{-}_i&=|\Downarrow_i\rangle \langle\Uparrow_i|,
\end{align}
with $\hat{T}^{x}_i$ and $\hat{T}^z_i$ as exciton pseudo-spin operators: $\hat{T}^x_i=\frac{1}{2}(\hat{T}^{+}_{i}+\hat{T}^{-}_{i})$ and $\hat{T}^{z}_{i}=\frac{1}{2}[\hat{T}^{+}_{i},\hat{T}^{-}_{i}]$. One finds that
\begin{align}
    J_x&=\frac{16t^2}{2U-3J}, \\
    h_z&=U-U'+\frac{4\Delta^2}{J}.
\end{align}
The resulting effective Ising Hamiltonian in Eq.~(\ref{effH2}) captures the $Z_2$ charge-parity symmetry for orbitals. Nearest-neighbor pseudo-spin exchange follows from virtual tunneling processes between neighboring sites. Conversely, \textit{both} the crystal field splitting $\Delta$ and the difference between on-site and inter-orbital density interactions $(U-U')$ induce a transverse field that serves to polarize the pseudo-spins. The Ising transition in the pseudo-spin model at $J_x=h_z$ thus signals an onset of the EDW order with broken $Z_2$ symmetry.

\subsection{DQMC algorithm}

We characterize the Hubbard-Kanamori Hamiltonian in Eq.~(\ref{hamiltonian}) using DQMC. Here we present a brief introduction to the DQMC algorithm. Detailed introductions can be found in Refs.~\cite{PhysRevD.24.2278,PhysRevB.40.506,santos2003introduction}. 

In general, we would like to use DQMC to numerically calculate the expectation value of an observable $\hat{A}$ at an inverse temperature $\beta$:
\begin{align}
    \langle\hat{A}\rangle=\frac{\mathrm{tr}(\hat{A}e^{-\beta\hat{H}})}{\mathrm{tr}(e^{-\beta\hat{H}})}.
\label{observable}
\end{align}
To evaluate the trace in the presence of four-fermion interaction terms, we perform a Hubbard-Stratonovich (HS) transformation and decompose $\mathrm{e}^{-\beta\hat{H}}$ as a sum of HS field configuration dependent density matrices $\Tilde{\rho}_{\mathbf{s}}$: $\mathrm{e}^{-\beta\hat{H}}\propto \sum_{\{\mathbf{s}\}}\Tilde{\rho}_{\mathbf{s}}$, where $\{\mathbf{s}\}$ is the set of all possible HS field configurations in spacetime. Eq.~(\ref{observable}) is then re-expressed in a HS fields-dependent form as
\begin{align}
  \langle\hat{A}\rangle=\frac{\sum_{\{\mathbf{s}\}}\mathrm{tr}(\hat{A}\Tilde{\rho}_{\mathbf{s}})}{\sum_{\{\mathbf{s}\}}\mathrm{tr}(\Tilde{\rho}_{\mathbf{s}})}=\frac{\sum_{\{\mathbf{s}\}}\langle\hat{A}\rangle_{\mathbf{s}}w_{\mathbf{s}}}{\sum_{\{\mathbf{s}\}}w_{\mathbf{s}}},  
\end{align}
with $w_{\mathbf{s}}\coloneqq\mathrm{tr}(\Tilde{\rho}_{\mathbf{s}})$ and $\langle\hat{A}\rangle_{\mathbf{s}}\coloneqq\mathrm{tr}(\hat{A}\Tilde{\rho}_{\mathbf{s}})/\mathrm{tr}(\Tilde{\rho}_{\mathbf{s}})$. Thus, $w_{\mathbf{s}}$ can be naturally interpreted as the probability weight of the configuration $\mathbf{s}$. We have \cite{PhysRevD.24.2278,PhysRevB.40.506,santos2003introduction}
\begin{align}
    w_{\mathbf{s}}=\det \left[ \mathbf{I}+\mathbf{B}_{\mathbf{s}}\right],
    \label{det}
\end{align}
where $\mathbf{I}$ is the identity matrix and $\mathbf{B}_{\mathbf{s}}$ is a matrix which depends on the the HS configuration $\mathbf{s}$.

\subsection{Sign-Free Decomposition}

For the Hubbard-Kanamori model, the straightforward decomposition of the interaction terms leads to a severe sign problem immediately away from the sign-free special case in which all orbitals preserve particle-hole symmetry, overall and individually. Previously, several decomposition strategies were proposed to alleviate the sign problem in restricted parameter spaces \cite{motome1997quantum,motome1998numerical,PhysRevB.70.172504,PhysRevB.74.155102}; and more recently, Ref.~\cite{PhysRevB.99.235142} shows a general framework to control the sign problem with constrained path and phaseless approximations in the ground-state variant of DQMC. 
Here, we provide a sign-free decomposition strategy by exploiting a novel anti-unitary symmetry of the model at overall half-filling and in a specific interaction parameter regime. Our approach is thus numerical exact.

We re-write $\hat{H}_{I}$ in Eq.~(\ref{hamiltonian}) in a particle-hole symmetric, quadratic form
\begin{align}
\label{decomp}
\hat{H}_{I}&=\hat{H}^{D}_{I}+\hat{H}^{H}_{I}, \nonumber \\
\hat{H}^{D}_{I}&=\left[\frac{U+U'}{4}(\hat{n}_{A\uparrow}+\hat{n}_{A\downarrow}+\hat{n}_{B\uparrow}+\hat{n}_{B\downarrow}-2)^{2}\right. \nonumber\\
&\left.+\frac{U-U'}{4}(\hat{n}_{A\uparrow}+\hat{n}_{A\downarrow}-\hat{n}_{B\uparrow}-\hat{n}_{B\downarrow})^{2}\right], \nonumber\\
\hat{H}^{H}_{I}&=\frac{J}{2}(\hat{c}^{\dagger}_{A\uparrow}\hat{c}_{B\uparrow}+\hat{c}^{\dagger}_{A\downarrow}\hat{c}_{B\downarrow}+\hat{c}^{\dagger}_{B\uparrow}\hat{c}_{A\uparrow}+\hat{c}^{\dagger}_{B\downarrow}\hat{c}_{A\downarrow})^{2}.
\end{align}
Note that the chemical potential term is absorbed into $\hat{H}_I^D$, and the model is particle-hole symmetric in the absence of a chemical potential shift for arbitrary values of $U,U',J$, and $\Delta$, guaranteeing overall half filling. The density-density term $\hat{H}^{D}_{I}$ and the Hund's interaction term $\hat{H}^{H}_{I}$ commute with one another:
\begin{align}
    \left[\hat{H}^{D}_{I},\hat{H}^{H}_{I}\right]=0
\end{align}

We discretize the imaginary time interval $[0, \beta]$ into $M$ slices, $\beta=M\Delta\tau$, and use the Trotter-Suzuki scheme to write
\begin{align}
e^{-\beta\hat{H}}\approx \prod^{M-1}_{l=0}e^{-\Delta\tau\hat{H}_{0}}e^{-\Delta\tau\hat{H}^{D}_{I}}e^{-\Delta\tau\hat{H}^{H}_{I}}. \label{trotter}
\end{align}

The Hund's term $\hat{H}^{H}_{I}$ can be decomposed via discrete spin-1 Ising HS fields:
\begin{align}
    &e^{-\Delta\tau\hat{H}^{H}_{I}}=\sum_{s=-1,0,1}\gamma^{h}_{s}(a)e^{s\lambda^{h}(a)\hat{h}}, \label{dc}
\end{align}
where $\hat{h}=\hat{c}^{\dagger}_{A\uparrow}\hat{c}_{B\uparrow}+\hat{c}^{\dagger}_{A\downarrow}\hat{c}_{B\downarrow}+\hat{c}^{\dagger}_{B\uparrow}\hat{c}_{A\uparrow}+\hat{c}^{\dagger}_{B\downarrow}\hat{c}_{A\downarrow}$, and
\begin{align}
    \gamma_{0}^{h}(a)&=1-2\gamma_{1}^{h}(a), \label{g1}\\
    \gamma_{\pm 1}^{h}(a)&=\frac{1}{3+2e^{-a/2}+e^{-a}}, \label{g2}\\
    \lambda^{h}(a)&=\frac{1}{2}\mathrm{arccosh}\left[\frac{1}{2}\left(-1-2e^{a/2}+3e^{-2a}\right.\right. \nonumber \\
    &\qquad\qquad\qquad \left.\left.-e^{-a}+2e^{-5a/2}+e^{-3a}\right)\right], 
\end{align}
and $a=\Delta\tau J$. 

The density interaction term can be decomposed as in Ref.~\cite{PhysRevLett.124.077601}. Notice that $\hat{N}=\hat{n}_{A\uparrow}+\hat{n}_{A\downarrow}+\hat{n}_{B\uparrow}+\hat{n}_{B\downarrow}-2$ and $\hat{M}=\hat{n}_{A\uparrow}+\hat{n}_{A\downarrow}-\hat{n}_{B\uparrow}-\hat{n}_{B\downarrow}$ both take values in $\{0,\pm 1,\pm 2\}$, and the following relation holds for $x=0, \pm 1, \pm 2$:
\begin{align}
e^{-ax^{2}}=\sum_{s=-1,0,1}\gamma_{s}(a)e^{\lambda(a) sx}, 
\end{align}
where
\begin{align}
&\gamma_{0}(a)=\frac{1+\cosh(a)}{1+2\cosh(a)+\sinh(a)} \label{AF1}, \\
&\gamma_{\pm 1}(a)=\frac{1}{3+e^{-2a}+2e^{-a}} \label{AF2}, \\
&\lambda(a)=\mathrm{arccosh}[\frac{1}{2}(e^{-3a}+e^{-2a}+e^{-a}-1)]  \label{AF3}, 
\end{align}
and $a=\Delta\tau(U+U')/4$ for $\hat{N}$, $a=\Delta\tau(U-U')/4$ for $\hat{M}$.

Using Eqs.~(\ref{dc}) through (\ref{AF3}), the four-fermion terms in Eq.~(\ref{trotter}) can be expressed in terms of bilinear operators coupled to three spin-1 Ising HS fields per site $\mathbf{s}=(s^{h},s,\bar{s})$, such that
\begin{align}
\mathrm{tr}(e^{-\beta\hat{H}})=\sum_{\mathbf{s}}w_{\mathbf{s}},
\end{align}
where 
\begin{align}
w_{\mathbf{s}}=\mathrm{tr}&\left[\left(\prod_{i,m}\gamma^{h}_{im}\gamma_{im}\bar{\gamma}_{im}e^{-2s_{im}\lambda}\right) \right. \nonumber \\
&\left.\prod_{m}e^{-\Delta\tau\hat{H}_{0}}e^{\sum_{i}s^h_{im}\lambda^h\hat{h}_{i}+s_{im}\lambda\hat{N}_{i}+\bar{s}_{im}\bar{\lambda}\hat{M}_{i}}\right]. 
\end{align}
$s^h_{im},\lambda^h,\gamma^h_{im}$ are coupled to Hund's terms $\hat{h}_{i}$; $s_{im},\lambda,\gamma_{im}$ are coupled to density-like terms $\hat{N}_{i}$; $\bar{s}_{im},\bar{\lambda},\bar{\gamma}_{im}$ are coupled to magnetization-like terms $\hat{M}_{i}$. It can be shown by evaluating the trace over the Hilbert space that
\begin{align}
w_{\mathbf{s}}=p_{\mathbf{s}}\,\mathrm{det}[\mathbf{I}+\mathbf{B}_{\mathbf{s}}],
\end{align}
where
\begin{align}
&p_{\mathbf{s}}=\prod_{i,m}\gamma^h_{im}\gamma_{im}\bar{\gamma}_{im}e^{-2s_{im}\lambda}, \\
&\mathbf{B}_{\mathbf{s}}\equiv\prod_{m}e^{-\Delta\tau\mathbf{H}_{0}}e^{\sum_{i}s^h_{im}\lambda^h\mathbf{h}_{i}}e^{\sum_{i}s_{im}\lambda\mathbf{N}_{i}}e^{\sum_{i}\bar{s}_{im}\bar{\lambda}\mathbf{M}_{i}},
\end{align}
and $\mathbf{I}$ is the identity matrix. Bold letters represent matrices, which are related to the corresponding operators by $\mathbf{c}^{\dagger}\mathbf{O}\mathbf{c}=\hat{O}$, where $\mathbf{c}^{\dagger}=(..., \hat{c}^{\dagger}_{i,A,\uparrow}, \hat{c}^{\dagger}_{i,A,\downarrow}, \hat{c}^{\dagger}_{i,B,\uparrow}, \hat{c}^{\dagger}_{i,B,\downarrow}, ...)$ is a vector of fermionic operators. 

The prefactors $p_{\mathbf{s}}$ are always positive according to Eq.~(\ref{g1}), Eq.~(\ref{g2}), Eq.~(\ref{AF1}) and Eq.~(\ref{AF2}). One possible strategy to prove that the matrix part of the probability weights $w_{\mathbf{s}}$ are positive semi-definite is to show that $\mathbf{I}+\mathbf{B}_{\mathbf{s}}$ has an anti-unitary symmetry $\mathbf{T}$, i.e. $\mathbf{T}^2 = -\mathbf{I}$ and $\mathbf{T}^{-1}(\mathbf{I}+\mathbf{B}_{\mathbf{s}})\mathbf{T}=\mathbf{I}+\mathbf{B}_{\mathbf{s}}$\cite{koonin1997shell}. It can be shown that the sign-free parameter regime for the particle-hole symmetric two-orbital Hubbard-Kanamori model in Eq.~(\ref{decomp}) is
\begin{align}
    |U| \leq U',
\end{align}
and $J$ can be positive, negative, or zero. The nearest neighbor hopping is restricted to be intra-orbital, and the hopping amplitude is the same for the two orbitals. Appendix \ref{proof_of_sign} presents a more detailed proof. Therefore, unbiased, exact numerical simulations can be performed for this model. 

\section{Results}

We perform sign-free DQMC simulations for systems with linear size $L=12$ at inverse temperature $\beta=12/t$, and focus on the onset of excitonic density order in regimes where the inter-orbital interaction $U'$ is slightly larger than the onsite interaction $U$, promoting an effective attraction between electrons and holes on opposite orbitals. To this end, we set $U'=6t$ and $U=5t$, and investigate the competition of Hund's exchange $J$ and crystal field splitting $\Delta$. While $J/t > 0$ in most transition-metal compounds, inverted Hund's coupling $J/t < 0$ has been proposed as a minimal model to explain superconductivity in the fullerides \cite{nomura2015unified}. In addition, the sign-free parameter regime we propose includes the orbital rotational symmetric point $U'=U-2J$ for inverted Hund's coupling. We will therefore chart a phase diagram primarily for the inverted Hund's coupling case, and complement the study of excitonic density orders by presenting additional results for the positive Hund's coupling case. 

\begin{figure}
\includegraphics[width=0.9\columnwidth]{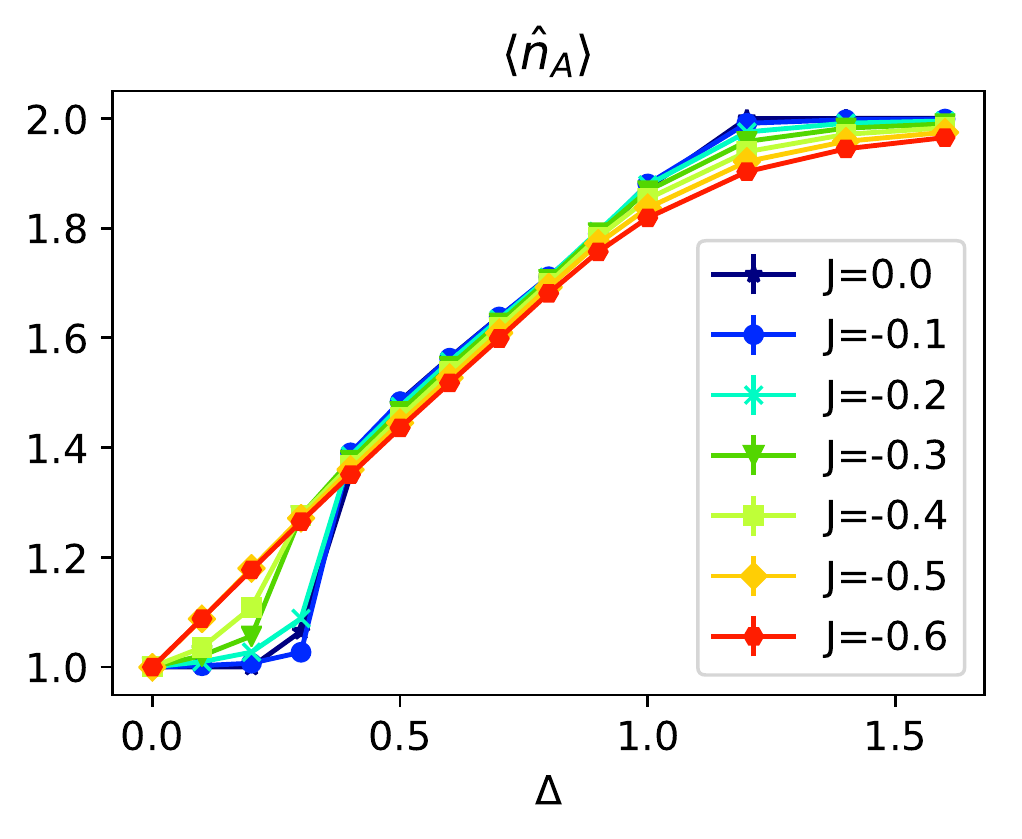}
\caption{\label{na} Electron number of orbital A averaged over sites and spins, with error bars for each data point taken from Monte Carlo estimates. Due to the sign-free nature of the DQMC algorithm we use this is work, the error bars are of the size of the marker for most of the data points in this plot and all the following plots.} 
\end{figure}

In the Hubbard-Kanamori model, the inclusion of a finite Hund's coupling $J$ introduces an additional source of ``strong coupling" compared to the standard single-orbital or single-band Hubbard model. Specifically, the interplay of Hund's coupling $J$ and crystal field $\Delta$ results in a variety of possible behaviours. While a negative $J$ tends to balance the electron number in the two orbitals and force spin anti-alignment, $\Delta$ lifts the orbital degeneracy and favors electron occupation of the lower energy orbital (orbital A with our definition). We focus on studying the competition of $J$ and $\Delta$ in the range $-0.6 \leq J/t \leq 0$ and $0 \leq \Delta/t \leq 1.5$. Figure~\ref{na} shows the average electron number per-site in orbital A, $\langle\hat{n}_{A}\rangle=\frac{1}{L^2}\sum_{i,\sigma}\langle\hat{n}_{iA\sigma}\rangle$, as a function of $J$ and $\Delta$. Since the Hamiltonian defined in Eq.~(\ref{decomp}) is particle-hole symmetric, the average electron number in orbital B is simply $\langle \hat{n}_{B}\rangle=2-\langle\hat{n}_{A}\rangle$. In the atomic limit at overall half-filling, the two electrons residing on a single site either separately occupy the two orbitals with spins anti-aligned, or both occupy the same orbital with lower energy (orbital A). Both cases present a ``low-spin" state in the atomic limit, and thus magnetic orders are generally not expected in our model. Notably, $\langle\hat{n}_{A}\rangle$ for $J=0\,t$ shows an edge at around $\Delta=0.4\,t$. This is a result of the energy gap in the charge sector caused by the checker board charge order formed for small $\Delta$. This order will be discussed in detail in the next section.

\subsection{Orbital and Spin Correlations}

We first study possible magnetic and orbital order, described via orbital $S_{o}(\vec{q})$ and spin $S_{s}(\vec{q})$ correlation functions
\begin{align}
    &S_{o}(\vec{q})=\frac{1}{L^{2}}\sum_{\vec{R},\vec{r}}e^{-i\vec{q}\cdot\vec{R}}\langle\hat{O}^{z}_{\vec{R}+\vec{r}}\hat{O}^{z}_{\vec{r}}\rangle,  \\
    &S_{s}(\vec{q})=\frac{1}{L^{2}}\sum_{\vec{R},\vec{r}}e^{-i\vec{q}\cdot\vec{R}}\langle\hat{S}^{z}_{\vec{R}+\vec{r}}\hat{S}^{z}_{\vec{r}}\rangle.  
\end{align}
Here, $\hat{O}^{z}_{\vec{r}}$ and $\hat{S}^{z}_{\vec{r}}$ represent the local spin and orbital polarization in the $z$-direction respectively:
\begin{align}
    &\hat{O}^{z}_{\vec{r}}=\sum_{\sigma}(\hat{n}_{\vec{r}A\sigma} - \hat{n}_{\vec{r}B\sigma}). \\
    &\hat{S}^{z}_{\vec{r}}=\sum_{\alpha}(\hat{n}_{\vec{r}\alpha\uparrow} - \hat{n}_{\vec{r}\alpha\downarrow}).
\end{align}
While the Hamiltonian is symmetric under SU(2) spin rotations, the combination of the pair hopping term of Hund's coupling and crystal field splitting retain only a $Z_2$ orbital symmetry that reflects the orbital charge parity.

\begin{figure}
\includegraphics[width=0.9\columnwidth]{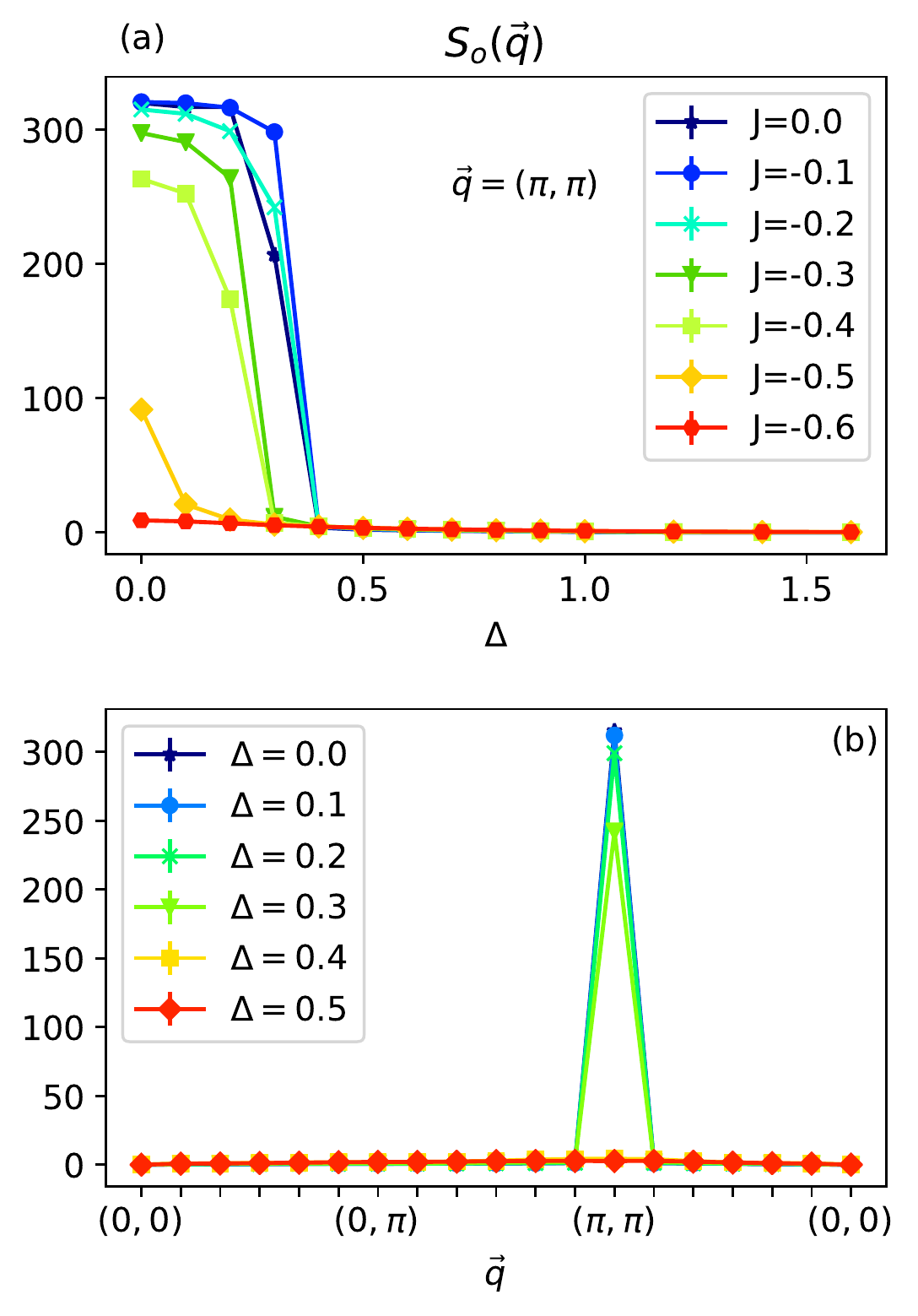}
\caption{\label{orbital} (a) $S_{o}(\pi,\pi)$ as a function of Hund's coupling $J$ and crystal field $\Delta$. (b) $S_{o}(\pi,\pi)$ for $J=-0.2t$ plotted along the high symmetry cuts in the first Brillouin zone.   }
\end{figure} 

To gain some intuition about orbital and magnetic ordering tendencies in the Hubbard-Kanamori model, 
first consider two limiting cases when $J=0\,t$ and both orbitals are at half-filling ($\Delta=0\,t$). In the limit where $U\gg t,U'$, at zero temperature the two layers decouple with $(\pi,\pi)$ antiferromagnetic (AFM) order. Conversely, large inter-orbital repulsion with $U'\gg t,U$ enforces an out-of-phase checkerboard charge order between the orbitals, which is termed anti-ferro-orbital (AFO) order. Away from these limits, especially at intermediate interaction strengths $U\approx U'$, numerical studies are essential to gain further insight. Reference~\cite{PhysRevLett.124.077601} shows that for finite $\Delta$ and $U' > U$, AFO order is gradually suppressed and eventually disappears at $\Delta \sim 0.4\,t$, giving way to a bi-excitonic condensate. In contrast, AFM order is absent at all values of $\Delta$ studied.

For finite $J$,  Hund's coupling competes with inter-orbital interactions and the crystal field. For $U' > U$, magnetic order remains suppressed over all values of $J$, with $S_{s}(\vec{q})$ uniform in momentum space and no discernible tendency towards AFM order. In contrast, the peak of the orbital correlation function $S_{o}(\vec{q})$ at $\vec{q}=(\pi,\pi)$ persists even for finite $J$. Figure~\ref{orbital}(a) shows the $\Delta$ and $J$ dependency of $S_{o}(\pi,\pi)$, and Fig.~\ref{orbital}(b) shows a set of representative results at $J=-0.2\,t$ plotted along the high symmetry cuts in momentum space. In the small $|J|$ and small $\Delta$ region, we find a sharp peak at $(\pi,\pi)$, which signals AFO order. The checkerboard AFO order preserves half-filling for both orbitals individually, and is gapped in the charge sector.

As $\Delta$ increases at fixed $J$ [Fig.~\ref{orbital}(a)], the crystal field gradually overcomes the charge gap and suppresses the peak in $S_{o}(\pi,\pi)$. This behavior also is evident from $\langle\hat{n}_{A}\rangle$, shown in Fig.~\ref{na} as a function of $\Delta$, and the system remains incompressible until the $S_{o}(\pi,\pi)$ peak is suppressed by $\Delta$.

\subsection{Excitonic Density Correlations}

\begin{figure}
\includegraphics[width=0.9\columnwidth]{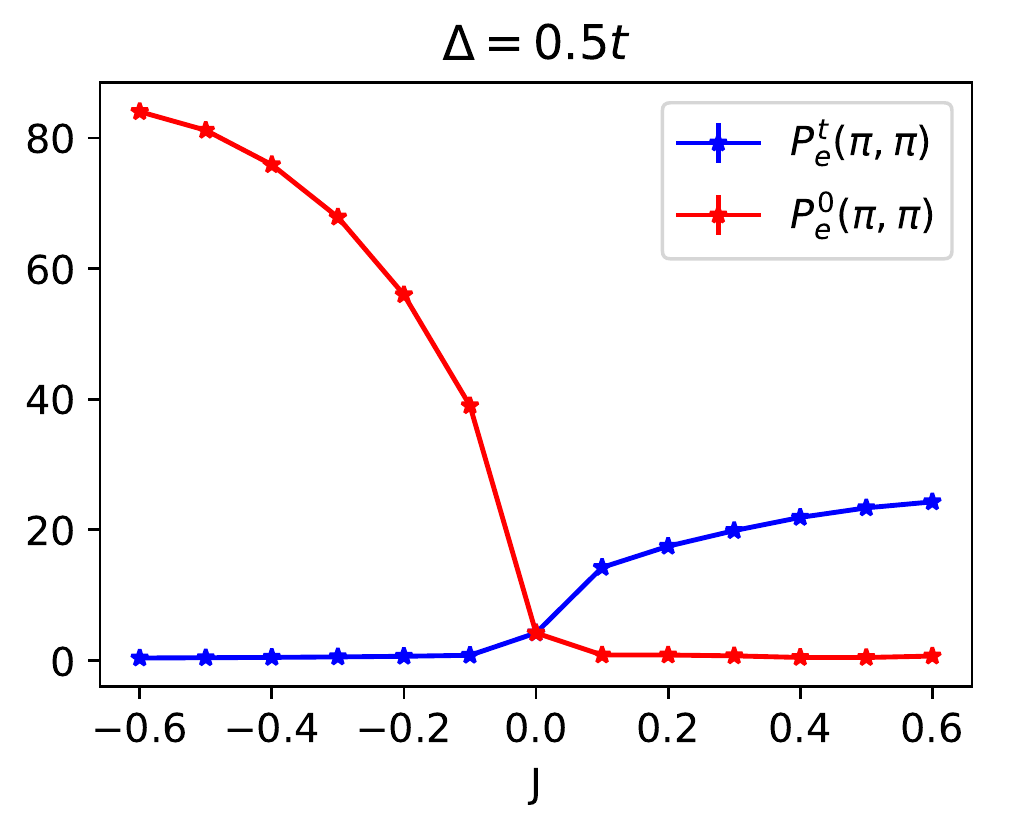}
\caption{\label{pjnj} Spin-singlet excitonic correlation $P^{0}_{e}(\pi,\pi)$ and spin-triplet excitonic correlation $P^{t}_{e}(\pi,\pi)$ measured for $\Delta=0.5t$ and $J$ ranging from $-0.6t$ to $0.6t$. The measurements are performed for a system with linear size $L=12$ at inverse temperature $\beta=12/t$, same as with the $J$-$\Delta$ results.}
\end{figure} 

We now turn to signatures of exciton formation in the two-orbital models. To this end, consider the inter-orbital spin-singlet and spin-triplet exciton creation operators\cite{PhysRevB.90.245144} 
\begin{align}
    &\hat{A}^{0\dagger}_{\vec{r}}=\frac{1}{\sqrt{2}}\sum_{\sigma}\hat{c}^{\dagger}_{\vec{r}A\sigma}\hat{c}_{\vec{r}B\sigma}\label{A0}, \\
    &\mathbf{\hat{A}}^{t\dagger}_{\vec{r}}=\frac{1}{\sqrt{2}}\sum_{\sigma\sigma'}\hat{c}^{\dagger}_{\vec{r}A\sigma}\mathbf{\sigma}_{\sigma\sigma'}\hat{c}_{\vec{r}B\sigma'}\label{A1},
\end{align}
where $\mathbf{\sigma}$ are Pauli matrices for spin. For simplicity, for spin-triplet excitons, we only consider the $x$ component of $\mathbf{\hat{A}}^{t\dagger}_{\vec{r}}$, which creates a spin anti-parallel exciton as $\hat{A}^{tx\dagger}_{\vec{r}}=\frac{1}{\sqrt{2}}(\hat{c}^{\dagger}_{\vec{r}A\uparrow}\hat{c}_{\vec{r}B\downarrow}+\hat{c}^{\dagger}_{\vec{r}A\downarrow}\hat{c}_{\vec{r}B\uparrow})$. This permits a straightforward definition of the corresponding correlation functions
\begin{align}
    &P_{e}^{0}(\vec{q})=\frac{1}{L^{2}}\sum_{\vec{R},\vec{r}}e^{-i\vec{q}\cdot\vec{R}}\langle(\hat{A}^{0\dagger}_{\vec{R}+\vec{r}}+\hat{A}^{0}_{\vec{R}+\vec{r}})(\hat{A}^{0\dagger}_{\vec{r}}+\hat{A}^{0}_{\vec{r}})\rangle, \\
    &P_{e}^{t}(\vec{q})=\frac{1}{L^{2}}\sum_{\vec{R},\vec{r}}e^{-i\vec{q}\cdot\vec{R}}\langle(\hat{A}^{tx\dagger}_{\vec{R}+\vec{r}}+\hat{A}^{tx}_{\vec{R}+\vec{r}})(\hat{A}^{tx\dagger}_{\vec{r}}+\hat{A}^{tx}_{\vec{r}})\rangle.
\end{align}
which will be used to characterize excitonic ordering tendencies. Notably, spin-singlet exciton condensation also preserves the global $SU(2)$ symmetry, permitting a finite-temperature phase transition. Conversely, spin-triplet exciton condensation remains stable in two dimensions only at zero temperature. 

To see whether singlet or triplet exciton formation is favored, notice that the spin-exchange part of the Hund's coupling in Eq.~(\ref{hamiltonian}) can be expressed using Eq.~(\ref{A0}) and Eq.~(\ref{A1}) \cite{PhysRevB.90.245144} as
\begin{align}
\mathbf{\hat{S}}_{iA}\cdot\mathbf{\hat{S}}_{iB}=-\frac{3}{4}\hat{A}^{s\dagger}\hat{A}^{s}+\frac{1}{4}\mathbf{\hat{A}}^{t\dagger}\mathbf{\hat{A}}^{t}.
\end{align}
Immediately, one observes that a spin anti-parallel triplet EDW becomes more energetically favourable for $J/t > 0$, whereas $J/t < 0$ favors spin-singlet EDWs. This phenomenon is confirmed in Fig.~\ref{pjnj}, which compares $P^{0}_{e}(\pi,\pi)$ and $P^{t}_{e}(\pi,\pi)$ at $\Delta=0.5\,t$ as a function of $J$. Notably, the peak of the singlet exciton correlation function for negative $J$ is significantly more pronounced than the peak of the triplet exciton correlation function for positive $J$. This is a direct consequence of discrete (continuous) symmetry breaking for spin-singlet (spin-triplet) EDWs in two dimensions, permitting (precluding) a finite-temperature phase transition.

\begin{figure}
\includegraphics[width=0.9\columnwidth]{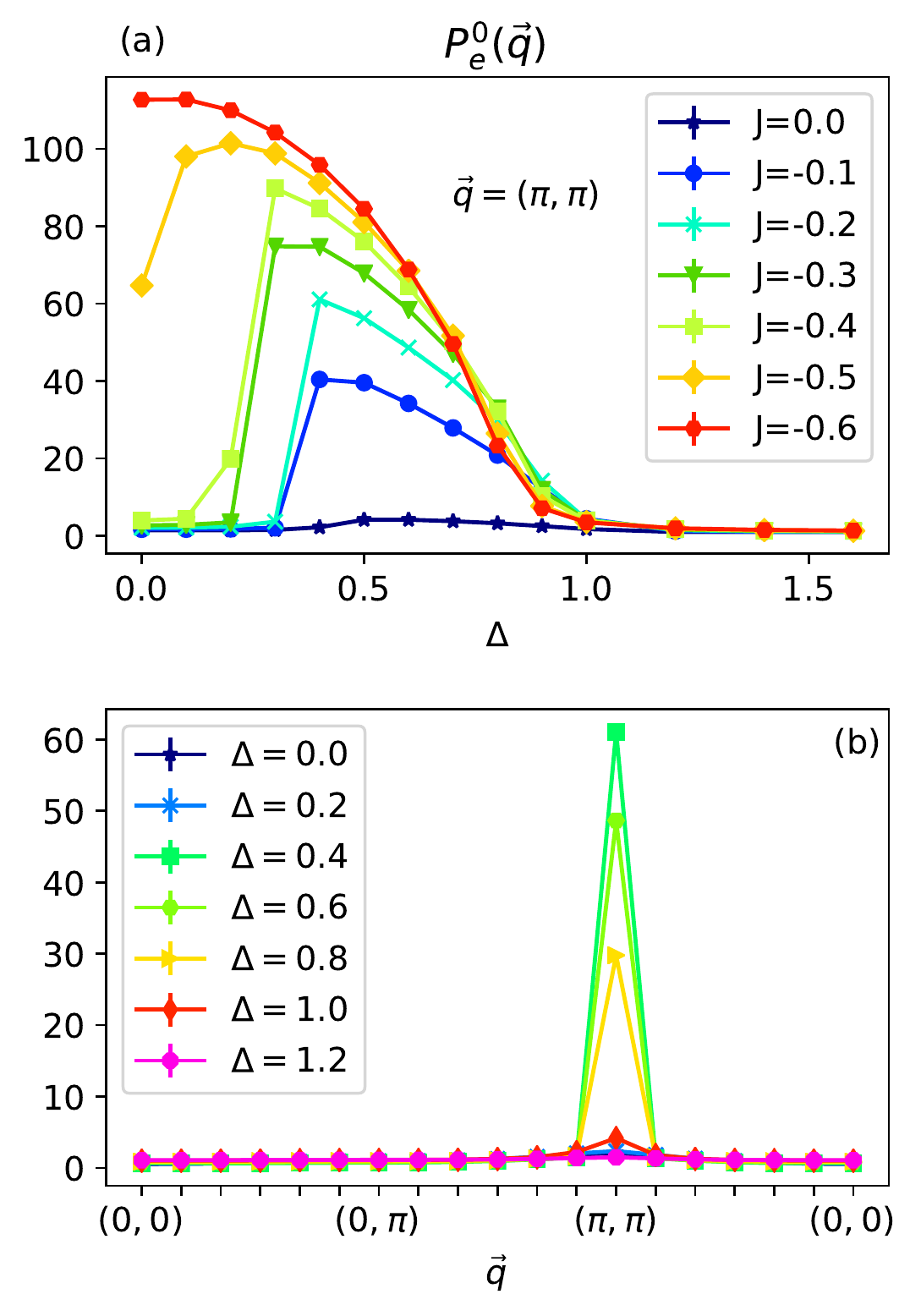}
\caption{\label{exc} (a) $P^{0}_{e}(\pi,\pi)$ as a function of Hund's coupling $J$ and crystal field $\Delta$. (b) $P^{0}_{e}(\pi,\pi)$ for $J=-0.2t$ plotted along the high symmetry cuts in the first Brillouin zone.}
\end{figure} 

\begin{figure}
\includegraphics[width=0.9\columnwidth]{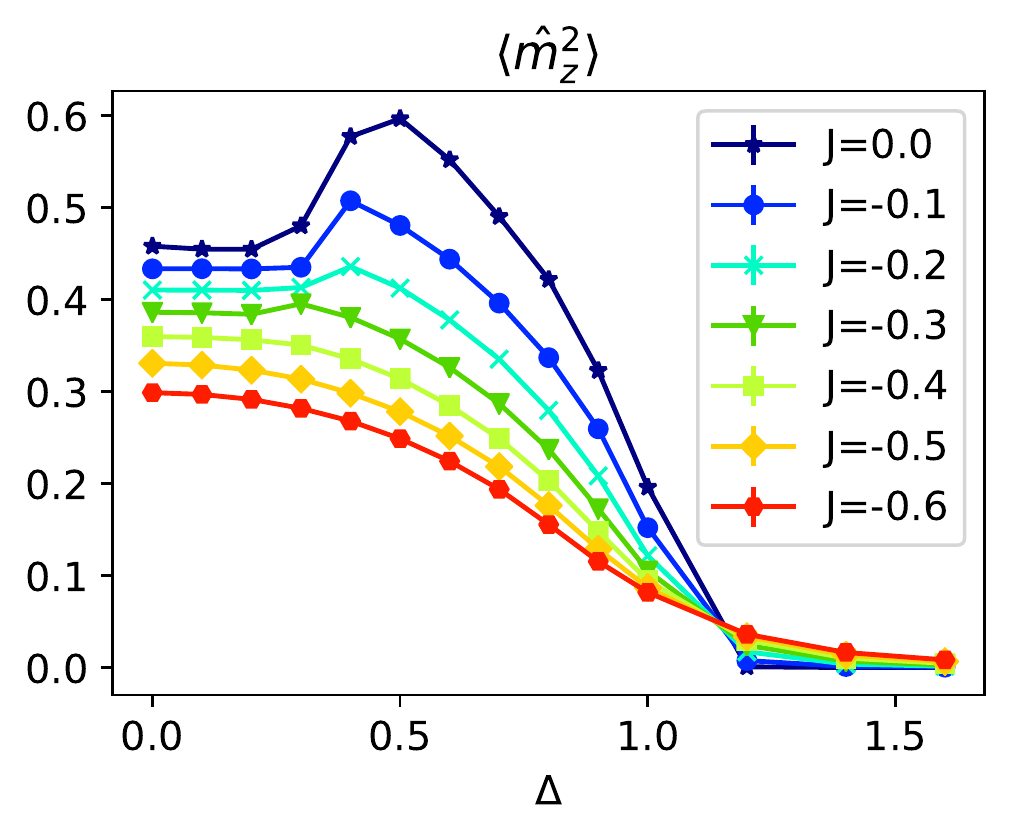}
\caption{\label{mz2} Local magnetic moment $\langle \hat{m}^{2}_{z}\rangle$ measured for various Hund's coupling $J$ and crystal field $\Delta$. }
\end{figure} 

\subsubsection{Competition with AFO}

For $J/t \leq 0$, a spin-singlet EDW competes with AFO. Figure~\ref{exc}(a) presents the $\Delta$ and $J$ dependency of $P^{0}_{e}(\pi,\pi)$, and Fig.~\ref{exc}(b) depicts a representative momentum space cut of $P^{0}_{e}(\vec{q})$ for $J=-0.2\,t$, showing that the peak of $P^{0}_{e}(\vec{q})$ always appears at $\vec{q}=(\pi,\pi)$. We start from an analysis of the $J=0\,t$ case. In the small $\Delta$ limit, the orbital $(\pi,\pi)$ modulation dominates the system as shown in Fig.~\ref{orbital}. On each site, one of the orbitals is doubly occupied, leaving the other orbital completely empty. This configuration efficiently blocks the orbital compensation and suppresses the inter-orbital electron-hole pairing. We obtain a $P^{0}_{e}(\vec{q})$ almost uniform in momentum space for this parameter region. As $\Delta$ increases, the excess of electrons (holes) in orbital A (B) increases as well, allowing higher pairing probability. $P^{0}_{e}(\pi,\pi)$ reaches the highest value at around $\Delta=0.5\,t$. Further increase of $\Delta$ completely fills orbital A (B) with electrons (holes) and a band insulating (BI) state is formed, eliminating both the orbital and excitonic $(\pi,\pi)$ modulation. The physical picture for $J/t < 0$ is an extension for the $J=0\,t$ case - the evolution from AFO to EC/EDW and finally to BI as $\Delta$ increases. However, the crossover value of $\Delta$ from AFO to EDW decreases as $|J|$ increases. Beyond $|J|=0.5\,t$, the $(\pi,\pi)$ peak for $S_{o}$ is absent even at $\Delta=0$, as shown in Fig.~\ref{orbital}. 

\subsubsection{Effect on Local Magnetic Moment}

The development of excitonic density modulation also has an effect on the local magnetic moment. Figure~\ref{mz2} depicts the magnetic moment averaged over all sites $\langle\hat{m}^{2}_{z}\rangle=\frac{1}{L^{2}}\sum_{i}\langle\hat{S}^{z2}_{i}\rangle$. We find that inverted Hund's coupling lowers $\langle\hat{m}^{2}_{z}\rangle$ for all values of the crystal field $\Delta$, as spin anti-alignment between orbitals is expected to be more energetically favorable in this case. For a specific $J$, the evolution of $\langle\hat{m}^{2}_{z}\rangle$ as a function of $\Delta$ closely follows the results of $P^{0}_{e}(\pi,\pi)$. This can be understood easily be considering the particle distribution among the two orbitals when the spin-singlet excitons are formed. In the strong $P^0_e(\pi, \pi)$ region, for a single site, the electron in orbital A pair up with the hole in orbital B, suppressing double occupation of the same orbital, which enhances the local magnetic moment for the two orbitals separately. This scenario can be further confirmed by studying the local magnetic moment for orbital A, defined as $\langle\hat{m}^{2}_{Az}\rangle=\frac{1}{L^{2}}\sum_{i}\langle(\hat{n}_{iA\uparrow}-\hat{n}_{iA\downarrow})^2\rangle$. Figure~\ref{mzmz}(a) shows the results for $\langle\hat{m}^{2}_{Az}\rangle$, which is consistent with the expectation. Appendix \ref{mzplots} presents more details regarding the magnetic moments. Nevertheless, the absolute value of $\langle\hat{m}^{2}_{z}\rangle$ remains small for $U' > U$ and no AFM order is present in the parameter region we study.

\begin{figure}
\includegraphics[width=0.9\columnwidth]{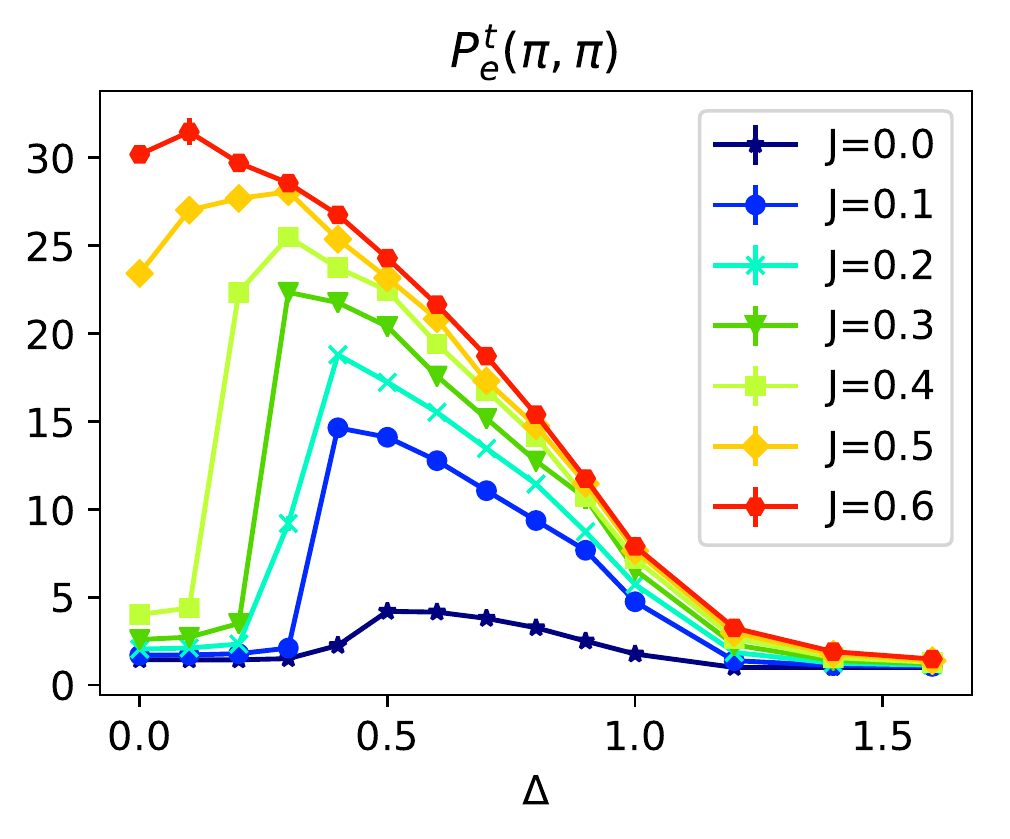}
\caption{\label{negj} Spin-triplet excitonic correlation function $P_e^t(\pi,\pi)$ measured for various positive Hund's coupling $J$ and crystal field $\Delta$. }
\end{figure} 

\subsubsection{Positive $J$ Regime}

For the positive Hund's coupling $J/t > 0$ case, we focus on presenting the results for spin-triplet EDW, which is energetically more favourable than its spin-singlet counterpart. Figure~\ref{negj} shows the spin-triplet excitonic correlation function $P_e^t(\pi,\pi)$ for positve Hund's coupling $J$ and various crystal field values $\Delta$. Interestingly, the spin-triplet results resemble the behavior of the spin-singlet excitonic correlation function in the negative $J$ regime.

\begin{figure}
\includegraphics[width=0.9\columnwidth]{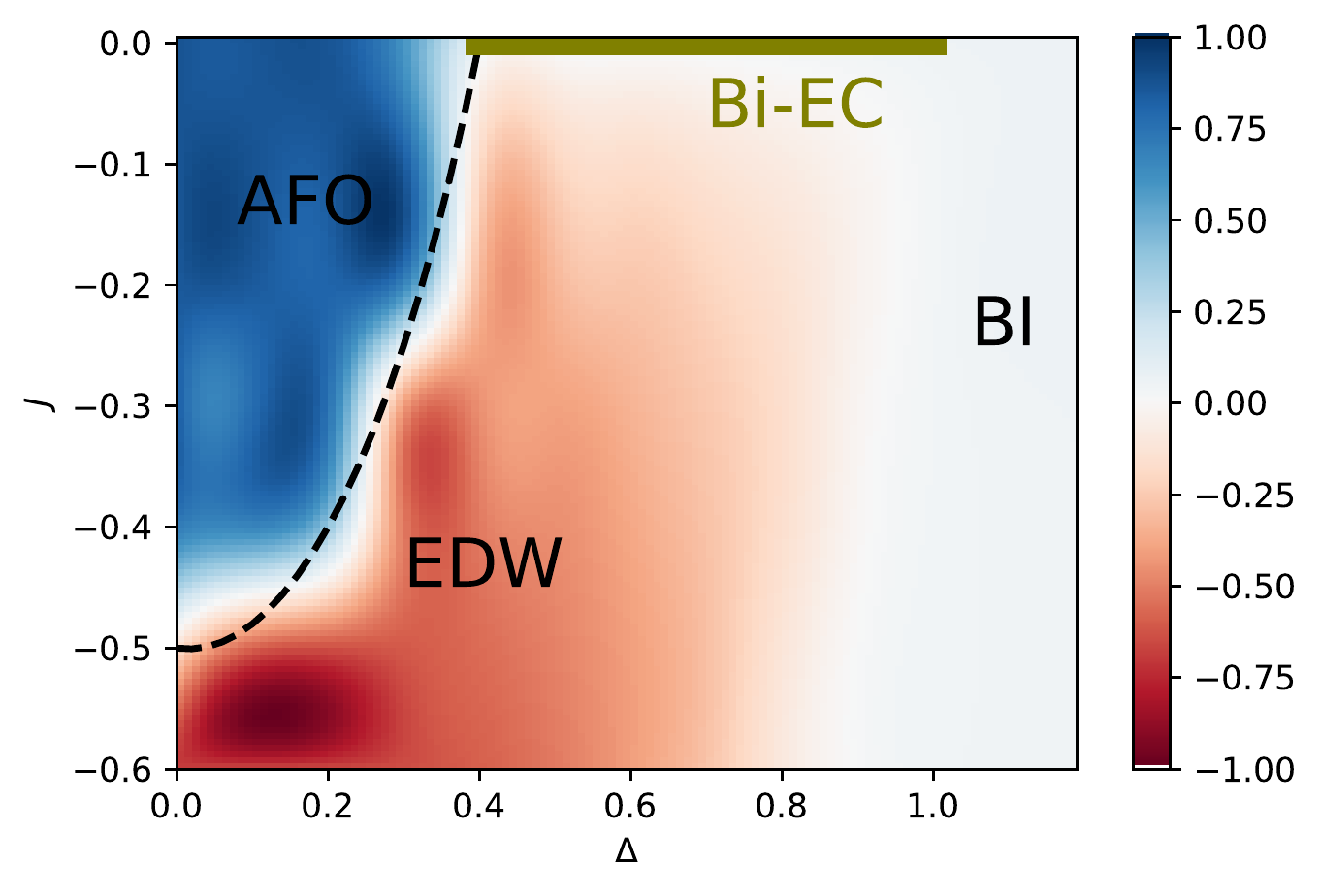}
\caption{\label{phased} Schematic plot of orders showing dominant correlations in the parameter regime $-0.6t\leq J\leq 0t$, $0t\leq \Delta\leq 1.6t$. The pseudo-color displayed is based on a normalized interpolation of the orbital correlation function $S_{o}(\pi,\pi)$ in Fig.~\ref{orbital}(a) and the spin-singlet excitonic correlation function $P^{0}_{e}(\pi,\pi)$ in Fig.~\ref{exc}(a). The bi-excitonic condensation (Bi-EC) phase locates at $J=0t$. The parameter range of Bi-EC is taken from Ref.~\cite{PhysRevLett.124.077601}, where the condensation phase if confirmed with a finite size analysis. }
\end{figure} 

\section{Conclusions}
In summary, we identify Hund's exchange-induced EDW order in a two-orbital Hubbard-Kanamori model, using numerically unbiased, fermion sign-problem-free DQMC simulations. Figure~\ref{phased} summarizes the various ordering tendencies, highlighting the dominant correlations as a function of Hund's exchange for $-0.6 \leq J/t \leq 0$ and orbital splitting $0 \leq \Delta/t \leq 1.6$. The false color is constructed by interpolating $S_{o}(\pi,\pi)$ and $P^{0}_{e}(\pi,\pi)$ results in Fig.~\ref{orbital}(a) and Fig.~\ref{exc}(a) after normalization, and shows the competition between AFO and EDW correlations. In the inverted Hund's coupling regime, $J$ favors balanced electron numbers in the two orbitals and spin anti-alignment, enhancing spin-singlet EDW correlations. Conversely, $\Delta$ polarizes the orbitals and drives the system towards a BI state in the high $\Delta$ limit. Here, strong inverted Hund's interactions permit a particularly simple description of the onset of EDW order, whereby the low energy behavior maps onto an effective exciton pseudo-spin Ising model in a transverse field, which succinctly describes the $Z_2$-breaking EDW transition. For positive Hund's coupling, a spin-triplet EDW becomes energetically favorable, displaying an analogous competition with AFO order. 

The DQMC algorithm we present in this work permits a sign-problem-free numerical simulation for a broad parameter range with $|U|\leq U'$, and $J$ of arbitrary sign. In particular, the sign-problem-free parameter range covers the rotationally invariant regime $U'=U-2J$ for inverted Hund's coupling. This enables extensions of our work to systematically explore and understand the emergence of symmetry breaking states, including unconventional superconductivity, in alkali-doped fullerides where the Hund's coupling is negative and the inter-orbital interaction can be larger than the intra-orbital interaction due to Jahn-Teller coupling\cite{nomura2015unified}. In addition, a two-orbital Hubbard-Kanamori model, like the one presented here, may be relevant for \ce{V_2O_3}, where the $V^{3+}$ $t_{2g}$ ($e^{\pi}_g$) orbitals near the Fermi energy are occupied by two electrons, and correlations induce an enhancement of the crystal-field splitting~\cite{PhysRevB.76.085127,PhysRevB.86.155121}. Local structural properties and the effective electron doping level of \ce{V_2O_3} can be altered and fine-tuned by Cr or Ti substitution~\cite{PhysRevLett.121.106401}, which allows a potential realization of the EDW orders proposed in our work.

\begin{acknowledgments}
X.X.H., B.M. and T.P.D. acknowledge support from the US Department of Energy, Office of Science, Office of Basic Energy Sciences, Division of Materials Sciences and Engineering, under Contract No. DE-AC02-76SF00515. M.C. acknowledges support from NSF EAGER 2132591.
\end{acknowledgments}

\appendix
\section{Proof of the Sign-Free Parameter Regime}
\label{proof_of_sign}
\begin{figure}
\includegraphics[width=0.9\columnwidth]{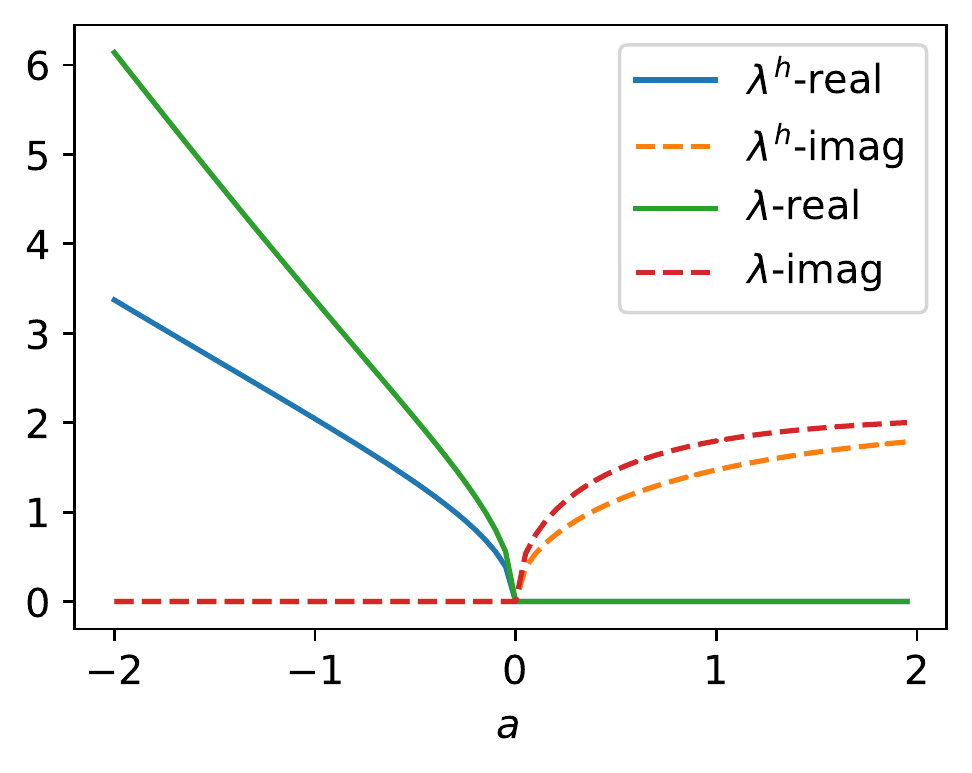}
\caption{\label{lambda} Discrete spin-1 decoupling parameters for the interaction terms. Both $\lambda$ and $\lambda^{h}$ are purely imaginary when $a \geq 0$ and purely real when $a\leq 0$.}
\end{figure}

The decomposition scheme is sign-free if there exists an anti-unitary symmetry $\mathbf{T}$, which satisfies $\mathbf{T}^{2}=-\mathbf{I}$, and is preserved by the following terms: $\Delta\hat{M}$, $\lambda^{h}\hat{h}$, $\lambda\hat{N}$, and $\bar{\lambda}\hat{M}$. Under the transformation $\mathbf{T}$, the coefficients ($\Delta,\lambda^{h},\lambda,\bar{\lambda}$) become their complex conjugates. Thus they are even under such a transformation if they are purely real and odd if they are purely imaginary. Since $\Delta$ denotes the electron-hole doping level, it is always purely real and even under the transformation $\mathbf{T}$. The real part and the imaginary part of the remaining coefficients ($\lambda^{h},\lambda,\bar{\lambda}$) depend on the parameters of the Hamiltonian and are shown in Fig.~\ref{lambda}. The variable $a$ in Fig.~\ref{lambda} takes the value $\Delta\tau J$, $\Delta\tau(U+U')/4$, and $\Delta\tau(U-U')/4$ for the three coefficients, respectively.

\begin{figure}
\includegraphics[width=0.9\columnwidth]{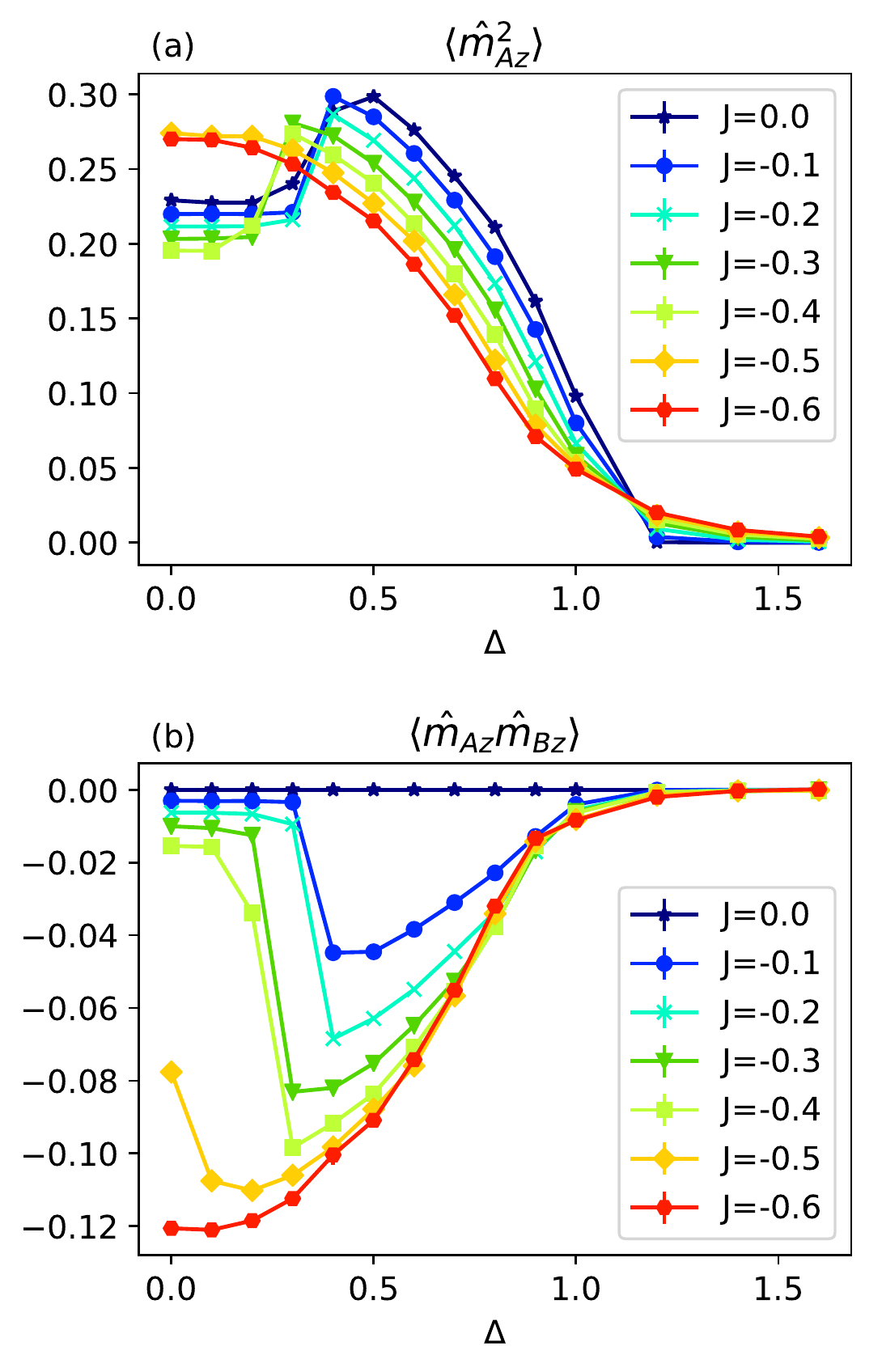}
\caption{\label{mzmz} (a) Local magnetic moment $\langle \hat{m}^{2}_{Az}\rangle$ (b) Inter-orbital magnetic moment cross-correlation $\langle \hat{m}_{Az}\hat{m}_{Bz}\rangle$ measured for various Hund's coupling $J$ and crystal field $\Delta$. }
\end{figure} 

\subsection{$J/t < 0$ Case}
We perform a single layer particle-hole transformation $\hat{c}_{iB\sigma}\rightarrow (-1)^{\delta_{i}}\hat{c}^{\dagger}_{iB\sigma}$ on layer B, where $\delta_{i}$ is even/odd on neighbouring sites. The kinetic energy term of the Hamiltonian is invariant under this transformation. $\hat{h}$ transforms as
\begin{align}
    \hat{h}\rightarrow (-1)^{\delta_{i}}(\hat{c}^{\dagger}_{A\uparrow}\hat{c}^{\dagger}_{B\uparrow}+\hat{c}_{B\uparrow}\hat{c}_{A\uparrow}+\hat{c}^{\dagger}_{A\downarrow}\hat{c}^{\dagger}_{B\downarrow}+\hat{c}_{B\downarrow}\hat{c}_{A\downarrow}), 
\end{align}
while $\hat{N}$ and $\hat{M}$ transform into each other
\begin{align}
    \hat{N}&\rightarrow\hat{M}, \\
    \hat{M}&\rightarrow\hat{N}. 
\end{align}
Consider an anti-unitary symmetry
\begin{align}
\hat{T} &= \sum_{i\sigma}[|i,A,\sigma\rangle\langle i,B,\sigma |-|i,B,\sigma\rangle\langle i,A,\sigma |]\hat{K}.
\end{align}
It is straightforward to verify that under this transformation, $(-1)^{\delta_{i}}(\hat{c}^{\dagger}_{A\uparrow}\hat{c}^{\dagger}_{B\uparrow}+\hat{c}_{B\uparrow}\hat{c}_{A\uparrow}+\hat{c}^{\dagger}_{A\downarrow}\hat{c}^{\dagger}_{B\downarrow}+\hat{c}_{B\downarrow}\hat{c}_{A\downarrow})$, $\hat{M}$, and $\hat{N}$ are even, odd, and even, respectively.
Therefore $\mathbf{I}+\mathbf{B}_{\mathbf{s}}$ is symmetric under $\mathbf{T}$ provided that $\lambda^{h}$ is purely real, $\lambda$ is purely imaginary, and $\bar{\lambda}$ is purely real. This corresponds to the parameter regime
\begin{align}
    |U|&\leq U' \\
    J/t&<0
\end{align}

\subsection{$J/t > 0$ Case}
We perform a particle-hole transformation for spin down operators $\hat{c}_{i\alpha\downarrow}\rightarrow (-1)^{\delta_{i}}\hat{c}^{\dagger}_{i\alpha\downarrow}$, where $\delta_{i}$ is even/odd on neighbouring sites. The kinetic energy term of the Hamiltonian is invariant under this transformation. $\hat{h}$, $\hat{N}$, and $\hat{M}$ transform as
\begin{align}
    \hat{h}&\rightarrow \hat{c}^{\dagger}_{A\uparrow}\hat{c}_{B\uparrow}+\hat{c}^{\dagger}_{B\uparrow}\hat{c}_{A\uparrow}-\hat{c}^{\dagger}_{A\downarrow}\hat{c}_{B\downarrow}-\hat{c}^{\dagger}_{B\downarrow}\hat{c}_{A\downarrow}, \\ 
    \hat{N}&\rightarrow \hat{n}_{A\uparrow}-\hat{n}_{A\downarrow}+\hat{n}_{B\uparrow}-\hat{n}_{B\downarrow}, \\
    \hat{M}&\rightarrow \hat{n}_{A\uparrow}-\hat{n}_{A\downarrow}-\hat{n}_{B\uparrow}+\hat{n}_{B\downarrow}.
\end{align}
Consider an anti-unitary symmetry
\begin{align}
\hat{T} &= \sum_{i}[|i,A,\uparrow\rangle\langle i,B,\downarrow |+|i,B,\downarrow\rangle\langle i,A,\uparrow |,\nonumber\\
&+|i,A,\downarrow\rangle\langle i,B,\uparrow |+|i,B,\uparrow\rangle\langle i,A,\downarrow |]\hat{K}.
\end{align}
It is straightforward to verify that under this transformation, the above terms are odd, odd, and even, respectively.
Therefore $\mathbf{I}+\mathbf{B}_{\mathbf{s}}$ is symmetric under $\mathbf{T}$ provided that $\lambda^{h}$ is purely imaginary, $\lambda$ is purely imaginary, and $\bar{\lambda}$ is purely real. This corresponds to the parameter regime
\begin{align}
    |U|&\leq U' \\
    J/t&>0
\end{align}

\section{Local Magnetic Moment for Inverted Hund's Coupling}
\label{mzplots}

Figure~\ref{mzmz}(a) shows the local magnetic moment measured for orbital A. The inter-orbital magnetic moment cross-correlation, 
defined as $\langle\hat{m}_{Az}\hat{m}_{Bz}\rangle=\frac{1}{L^{2}}\sum_{i}\langle\hat{n}_{iA\uparrow}-\hat{n}_{iA\downarrow})(\hat{n}_{iB\uparrow}-\hat{n}_{iB\downarrow})\rangle$, is plotted in Fig.\ref{mzmz}(b). For a fixed $J$, both $\langle\hat{m}^2_{Az}\rangle$ and the absolute value of $\langle\hat{m}_{Az}\hat{m}_{Bz}\rangle$ follow the results of $P^{0}_{e}(\pi,\pi)$, as a result of the particle distribution among the two orbitals enforced by inter-orbital particle-hole pairing. Meanwhile, inverted Hund's coupling favors spin anti-alignment between orbital A and orbital B, leading to a reduction in $\langle\hat{m}_{Az}\hat{m}_{Bz}\rangle$ as $|J|$ increases for a fixed $\Delta$. It also is straightforward to verify that $\langle\hat{m}^2_z\rangle=2\langle\hat{m}^2_{Az}\rangle+2\langle\hat{m}_{Az}\hat{m}_{Bz}\rangle$.

\bibliography{cite}

\end{document}